\pdfoutput=1
\documentclass[11pt,a4paper]{article}
\usepackage[hyperref]{emnlp2020}
\usepackage{times}
\usepackage{latexsym}
\usepackage{soul}
\usepackage{url}
\usepackage{graphicx}
\usepackage{amsmath}
\usepackage{booktabs}
\usepackage{latexsym}
\usepackage{microtype}
\usepackage{subfigure}
\usepackage{graphicx}
\usepackage{amssymb}
\usepackage[boxed,ruled,commentsnumbered]{algorithm2e}
\usepackage{multirow}
\usepackage{makecell}
\usepackage{verbatim}
\usepackage{changepage}

\usepackage{xcolor}
\newcommand{\bluefont}[1]{ #1}

\usepackage{microtype}

\aclfinalcopy 

\title{Privacy-Preserving News Recommendation Model Learning}

\author{Tao Qi$^1$, Fangzhao Wu$^2$, Chuhan Wu$^1$, Yongfeng Huang$^1$ and Xing Xie$^2$\\
  $^1$Department of Electronic Engineering \& BNRist, Tsinghua University, Beijing 100084, China  \\
  $^2$Microsoft Research Asia, Beijing 100080, China\\
  \tt{\{taoqi.qt,wuchuhan15\}@gmail.com~~yfhuang@tsinghua.edu.cn}\\
  \tt{\{fangzwu,xing.xie\}@microsoft.com} \\
  
  }

\date{}

\begin{document}
\maketitle

\begin{abstract}

News recommendation aims to display news articles to users based on their personal interest.
Existing news recommendation methods rely on centralized storage of user behavior data for model training, which may lead to privacy concerns and risks due to the privacy-sensitive nature of user behaviors.
In this paper, we propose a privacy-preserving method for news recommendation model training based on federated learning, where the user behavior data is locally stored on user devices. 
Our method can leverage the useful information in the behaviors of massive number users to train accurate news recommendation models and meanwhile remove the need of centralized storage of them.
More specifically, on each user device we keep a local copy of the news recommendation model, and compute gradients of the local model based on the user behaviors in this device.
The local gradients from a group of randomly selected users are uploaded to server, which are further aggregated to update the global model in the server.
Since the model gradients may contain some implicit private information, we apply local differential privacy (LDP) to them before uploading for better privacy protection.
The updated global model is then distributed to each user device for local model update.
We repeat this process for multiple rounds.
Extensive experiments on a real-world dataset show the effectiveness of our method in news recommendation model training with privacy protection.

\end{abstract}

\section{Introduction}

With the development of Internet and mobile Internet, online news websites and Apps such as Yahoo! News\footnote{https://news.yahoo.com} and Toutiao\footnote{https://www.toutiao.com/} have become very popular for people to obtain news information~\cite{okura2017embedding}.
Since massive news articles are posted online every day, users of online news services face heavy information overload~\cite{zheng2018drn}.
Different users usually prefer different news information.
Thus, personalized news recommendation, which aims to display news articles to users based on their personal interest, is a useful technique to improve user experience and has been widely used in many online news services~\cite{wu2019npa}.
The research of news recommendation has attracted many attentions from both academic and industrial fields~\cite{okura2017embedding,wang2018dkn,lian2018towards,an2019neural,wu2019ijcai}.

Many news recommendation methods have been proposed in recent years~\cite{wang2018dkn,wu2019npa,danzhu2019}.
These methods usually recommend news based on the matching between the news representation learned from news content and the user interest representation learned from historical user behaviors on news.
For example, Okura et al.~\shortcite{okura2017embedding} proposed to learn news representations from the content of news articles via autoencoder, and learn user interest representations from the clicked news articles via Gated Recurrent Unit (GRU) network.
They ranked the candidate news articles using the direct dot product of the news and user interest representations.
These approaches all rely on the centralized storage of user behavior data such as news click histories for model training.
However, users' behaviors on news websites and Apps are privacy-sensitive, the leakage of which may bring catastrophic consequences.
Unfortunately, the centralized storage of user behavior data in server may lead to high privacy concerns from users and risks of large-scale private data leakage.

In this paper, we propose a privacy-preserving method for news recommendation model training.
Instead of storing user behavior data on a central server, in our method it is locally stored on (and never leaves) users' personal devices, which can effectively reduce the privacy concerns and risks~\cite{mcmahan2017communication}.
Since the behavior data of a single user is insufficient for model training, we propose a federated learning based framework named \textit{FedNewsRec} to coordinate massive user devices to collaboratively learn an accurate news recommendation model without the need to centralized storage of user behavior data.
In our framework, on each user device we keep a local copy of the news recommendation model. 
Since the user behaviors on news websites or Apps stored in user device can provide important supervision information of how the current model performs, we compute the model gradients based on these local behaviors.
The local gradients from a group of randomly selected users are uploaded to server, which are further aggregated to update the global news recommendation model maintained in the server.
The updated global model is then distributed to each user device for local model update.
We repeat this process for multiple rounds until the training converges.
Since the local model gradients may also contain some implicit private information of users' behaviors on their devices, we apply the local differential privacy (LDP) technique to these local model gradients before uploading them to server, which can better protect user privacy at the cost of slight performance sacrifice.
We conduct extensive experiments on a real-world dataset.
The results show that our method can achieve satisfactory performance in news recommendation by coordinating massive users for model training, and at the same time can well protect user privacy.

The major contributions of this work include:

    (1) We propose a privacy-preserving method to train accurate news recommendation model by leveraging the behavior data of massive users and meanwhile remove the need to its centralized storage  to protect user privacy.
    
    (2) We propose to apply local differential privacy to protect the private information in local gradients communicated between user devices and server.
    
    (3) We conduct extensive experiments on a real-world dataset to verify the proposed method in recommendation accuracy and privacy protection.

\section{Related Work}

\subsection{News Recommendation}

News recommendation can be formulated as a problem of matching between news articles and users.
There are three core tasks for news recommendation, i.e., how to model the content of news articles (news representation), how to model the personal interest of users in news (user representation), and how to measure the relevance between news content and user interest.
For news representation, many feature based methods have been applied.
For example, Lian et al.~\shortcite{lian2018towards} represented news using URLs, categories and entities.
Recently, many deep learning based news recommendation methods represent news from the content using neural networks.
For example, Okura et al.~\shortcite{okura2017embedding} used denoising autoencoder to learn news representations from news content.
Wu et al.~\shortcite{wu2019neurald} proposed to learn news representations from news titles via multi-head self-attention network.
For user representation, existing news recommendation methods usually model user interest from their historical behaviors on news platforms.
For example, Okura et al.~\shortcite{okura2017embedding} learned user representations from the previously clicked news using GRU network.
An et al.~\shortcite{an2019neural} proposed a long- and short-term user representation model (LSTUR) for user interest modeling.
It captures the long-term user interest via user ID embedding and the short-term user interest from the latest news click behaviors via GRU.
For measuring the relevance between user interest and news content, dot product of user and news representation vectors is widely used~\cite{okura2017embedding,wu2019ijcai,an2019neural}.
Some methods also explore cosine similarity~\cite{danzhu2019}, feed-forward network~\cite{wang2018dkn}, feature-interaction network~\cite{lian2018towards}.

These existing news recommendation methods all rely on centrally-stored user behavior data for model training.
However, users' behaviors on news platforms are privacy-sensitive.
The centralized storage of user behavior data may lead to serious privacy concerns of users.
In addition, the news platforms have high responsibility to prevent user data leakage, and have high pressure to meet the requirements of strict user privacy protection regulations like GDPR\footnote{https://eugdpr.org/}.
Different from existing news recommendation methods, in our method the user behavior data is locally stored on personal devices, and only the model gradients are communicated between user devices and server.
Since the model gradients contain much less user information than the raw behavior data and they are futher processed by the Local Differential Privacy (LDP) technique, our method can protect user privacy much better than existing news recommendation methods.

\subsection{Federated Learning}

Federated learning~\cite{mcmahan2017communication} is a privacy-preserving machine learning technique which can leverage the rich data of massive users to train shared intelligent models without the need to centrally store the user data.
In federated learning the user data is locally stored on user mobile devices and never uploaded to server.
Instead, each user device computes a model update based on the local data, and the locally-computed model updates from many users are aggregated to update the shared model.
Since model updates usually contain much less information than the raw user data, the risks of privacy leakage can be effectively reduced.
Federated learning requires that the labeled data can be inferred from user interactions for supervised model learning, which can be perfectly satisfied in our news recommendation scenario, since the click and skip behaviors on news websites and Apps can provide rich supervision information.

Federated learning has been applied to training query suggestion model for smartphone keyboard and topic models~\cite{FedTM}.
There are also some explorations in applying federated learning to recommendation~\cite{ammad2019federated,chai2019secure}.
For example, Ammad et al.~\shortcite{ammad2019federated} proposed a federated collaborative filtering (FCF) method.
In FCF, the personal rating data is locally stored on user client and is used to compute the local gradients of user embeddings and item embeddings.
The user embeddings are locally maintained in user client and are directly updated using the local gradient on each client.
The item embeddings are maintained by a central server, and are updated using the aggregated gradients of many clients.
Chai et al.~\shortcite{chai2019secure} proposed a federated matrix factorization method, which is very similar with FCF.
However, these methods require all users to participate the process of federated learning to train their embeddings, which is not practical in real-world recommendation scenarios.
Besides, these methods represent items using their IDs, and are difficult to handle new items since many news articles are posted every day which are all new items.
Thus, these federated learning based recommendation methods have their inherent drawbacks, and are not suitable for news recommendation.

\subsection{Local Differential Privacy}

Local differential privacy (LDP) is an important technique to provide guarantees of privacy for sensitive information collection and analysis~\cite{ren2018textsf}.
It has attracted increasing attentions since user privacy protection has become a more and more important issue~\cite{kairouz2014extremal,qin2016heavy}.
A classical scenario of LDP is that there are a set of users, and each user $u$ has a private value $v$, which is sent to a untrusted third-party aggregator so that the aggregator can learn some statistical information of the private value distribution among the users~\cite{cormode2018privacy}.
LDP can guarantee that the leakage of private information for each individual user is bounded by applying a randomized algorithm $\mathcal{M}$ to private value $v$ and sending the perturbed value $\mathcal{M}(v)$ to the aggregator for statistical information inference.
The randomized algorithm $\mathcal{M}(\cdot)$ is called to satisfy $\epsilon$-local differential privacy if and only if for two arbitrary input private values $v$ and $v'$, the following inequation holds:
\begin{equation}
    \text{Pr}[\mathcal{M}(v) =y] \leq e^{\epsilon} \ \text{Pr}[\mathcal{M}(v') =y],
    \label{eq.ldp}
\end{equation}
where $y\in range(\mathcal{M})$.
$\epsilon \geq 0$, and it is usually called privacy budget.
Smaller $\epsilon$ means better private information protection.
In many works~\cite{sarathy2010some,duchi2013local}, $\mathcal{M}(\cdot)$ is implemented by adding Laplace noise to the private value.
In this paper we apply LDP technique to the model gradients which are generated in user devices based on user behaviors and uploaded to server, to better protect user privacy and remove the need to a trusted server.

\section{FedNewsRec for Privacy-Preserving News Recommendation}

In this section we introduce our \textit{FedNewsRec} method for privacy-preserving news recommendation model training.
We first describe the news recommendation model.
Then we describe the details of \textit{FedNewsRec}.

\subsection{Basic News Recommendation Model}

Following previous works~\cite{wu2019ijcai,an2019neural}, the news recommendation model in our method can be decomposed into two core sub-models, i.e., a \textit{news model} to learn news representations and a \textit{user model} to learn user representations.

\begin{figure}[t]
    \centering
    \resizebox{0.32\textwidth}{!}{
    \includegraphics{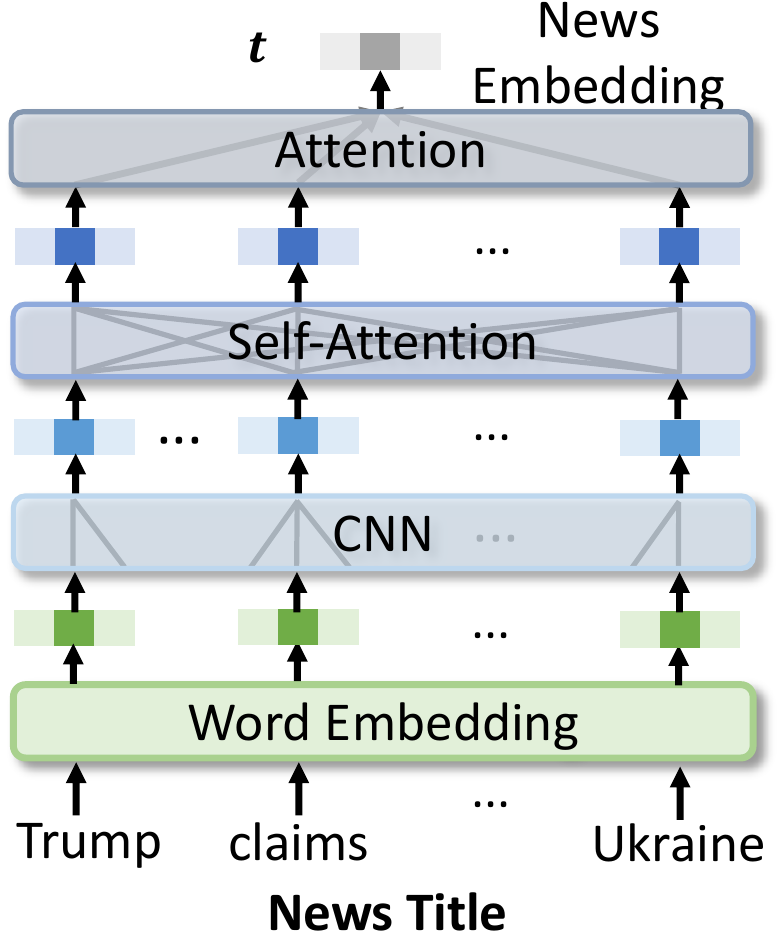}
    }
    \caption{The architecture of \textit{news model}.}
    \label{fig_news_encoder}
\end{figure}

\textbf{News Model} \ 
The \textit{news model} aims to learn news representations to model news content.
Its architecture is shown in Fig.~\ref{fig_news_encoder}.
Following~\cite{wu2019npa}, we learn news representations from news titles.
The \textit{news model} contains four layers stacked from bottom to up.
The first layer is word embedding, which converts the word sequence in a news title into a sequence of semantic word embedding vectors.
The second layer is a CNN network, which is used to learn word representations by capturing local contexts.
The third layer is a multi-head self-attention network~\cite{vaswani2017attention}, which can learn contextual word representations by modeling the long-range relatedness between different words.
The fourth layer is an attention network, which is used to build a news representation vector $\textbf{t}$ from the output of multi-head self-attention network by selecting informative words.





\begin{figure}[t]
    \centering
    \resizebox{0.48\textwidth}{!}{
    \includegraphics{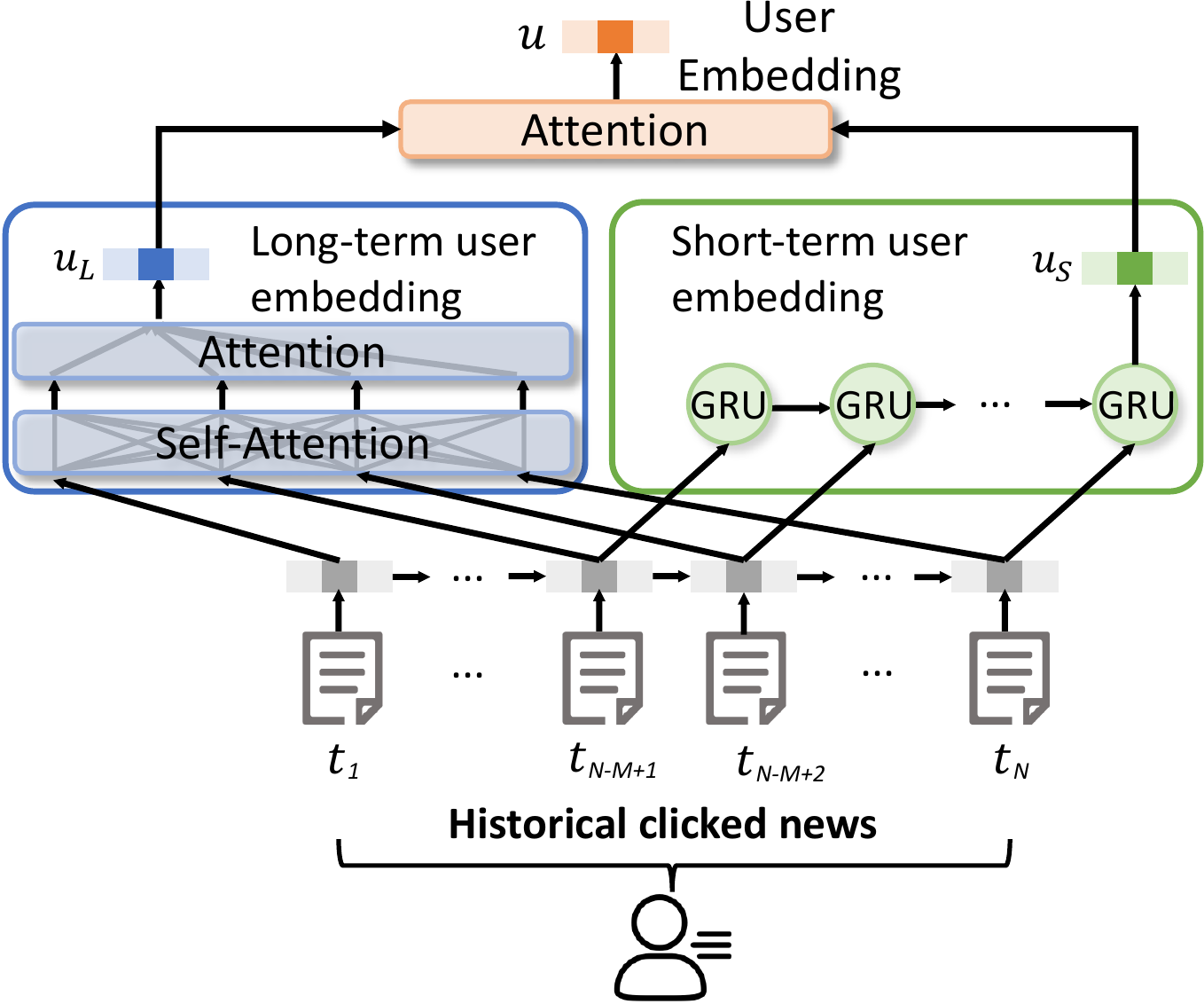}
    }
    \caption{The architecture of \textit{user model}.}
    \label{fig_user_encoder}
\end{figure}

\textbf{User Model} \ 
The \textit{user model} is used to learn user representations to model their personal interest.
Its architecture is shown in Fig.~\ref{fig_user_encoder}.
Following~\cite{okura2017embedding}, we learn user representation from their clicked news articles.
Motivated by the LSTUR model proposed by~\citet{an2019neural}, we learn representations of users by capturing both long-term and short-term interests.
The difference is that in LSTUR the embeddings of user IDs are used to model long-term interest, while in our \textit{user model} it is learned from all the historical behaviors through a combination of a multi-head self-attention network and an attentive pooling network.
This is because in the federated learning scenario, it is not practical that all users can participate the process of model training.
Thus, the ID embeddings of many users in LSUTR cannot be learned.
For short-term user interest modeling, our \textit{user model} applies a GRU network to the recent behaviors of users, which is the same with LSUTR.
The embeddings of long-term interest and short-term interest are combined with an attention network into a unified user embedding vector $\textbf{u}$.

\textbf{Model Training from User Behavior} \ 
Users' behaviors on news websites and Apps can provide useful supervision information to train the news recommendation models.
For example, if a user $u$ clicks a news article $t$ which has low ranking score predicted by the model, then we can tune the model to give higher ranking score for this user-news pair.
We propose to train the news recommendation model based on both click and non-click behaviors.
More specifically, following~\cite{wu2019npa}, for each news $t_i^c$ clicked by user $u$, we randomly sample $H$ news which are displayed in the same impression but not clicked.
Assume this user has $B_u$ click behaviors in total, then the loss function of the news recommendation model with parameter set $\Theta$ is defined as:
\begin{equation} 
    \mathcal{L}_u(\Theta) = \sum\limits_{i=1}^{B_u}  L^{i},
\end{equation} 
\begin{equation}
\label{eq.loss}
    L^{i} =  -\log(\frac{\exp(s(u, t_i^c))}{\exp(s(u, t_i^c)) + \sum_{j=1}^H \exp(s(u, t_{i,j}^{nc}))} ),
\end{equation}
where $t_i^c$ and $t_{i,j}^{nc}$ are clicked and non-clicked news articles shown in the same impression.
$s(u, t)$ is the ranking score of news $t$ for user $u$, which is defined as the dot product of their embedding vectors, i.e., $s(u, t)=\mathbf{u}^T\mathbf{t}$.

\begin{figure*}[t]
    \centering
     \resizebox{0.98\textwidth}{!}{
     \includegraphics{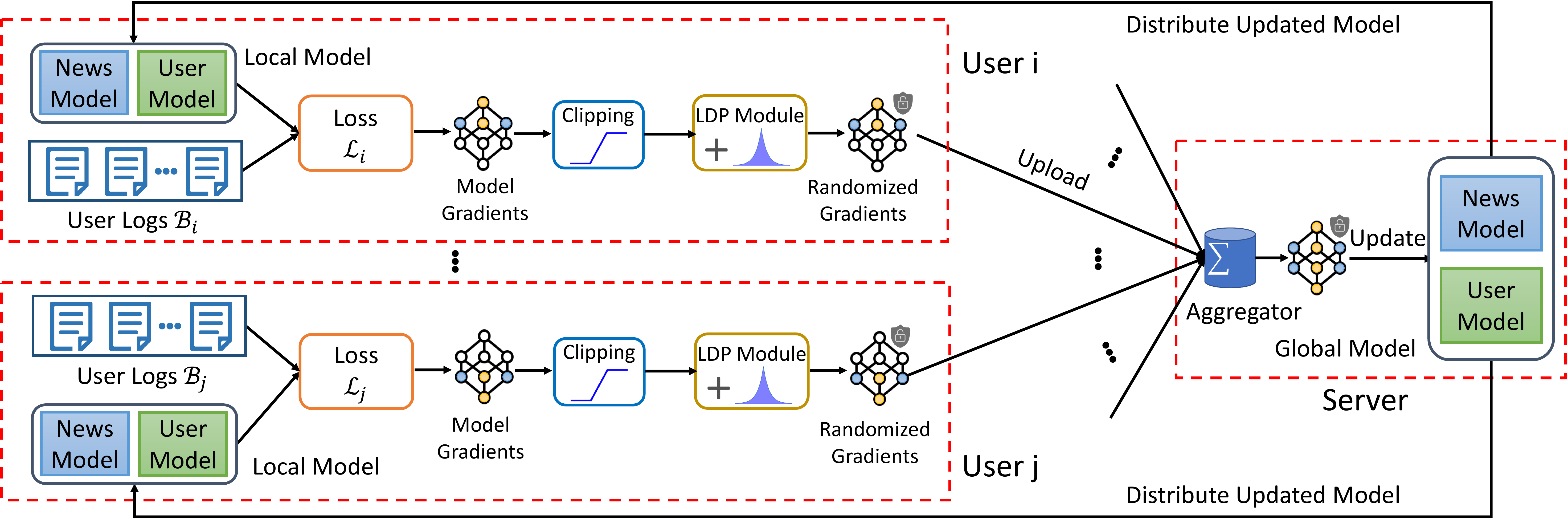}
     }
    \caption{The framework of our privacy-preserving news recommendation approach.}
    \label{fig_framework}
\end{figure*}

\subsection{The Framework of FedNewsRec}

Next, we introduce our \textit{FedNewsRec} framework for privacy-preserving news recommendation model training, which is shown in Fig.~\ref{fig_framework}. 
In our \textit{FedNewsRec} framework, the user behaviors on news platforms (websites or Apps) are locally stored on the user devices and never uploaded to server.
In addition, the servers which provide news services do not record nor collect the user behaviors, which can reduce the privacy concerns of users and the risks of data leakage.
Since an accurate news recommendation model can effectively improve users' news reading experiences and the behavior data from a single user is far from sufficient for training an accurate and unbiased model, in our \textit{FedNewsRec} framework we propose to coordinate a large number of user devices to collectively train intelligent news recommendation models. 

Following~\cite{mcmahan2017communication}, each user device which participates the model training is called a client.
Each client has a copy of the current news recommendation model $\Theta$ which is maintained by the server.
Assume user $u$'s client has accumulated a set of behaviors on news platforms which is denoted as $\mathcal{B}_u$, then we compute a local gradient of model $\Theta$ according to the behaviors $\mathcal{B}_u$ and the loss function defined in Eq.~(\ref{eq.loss}), which is denoted as $g_u=\frac{\partial \mathcal{L}_u }{\partial \Theta}$.
Although the local model gradient $g_u$ is computed from a set of behaviors rather than a single behavior, it may still contain some private information of user behaviors~\cite{deepleakge}.
Thus, for better privacy protection, we apply local differential privacy (LDP) technique to the local model gradients.
Denote the randomized algorithm applied to $g_u$ as $\mathcal{M}$, which is defined as:
\begin{equation}\label{eq.M}
    \mathcal{M}(g_u) = \text{clip}(g_u, \delta) + n,
\end{equation}
\begin{equation}\label{eq.La}
    n \sim La(0,\lambda),
\end{equation}
where $n$ is Laplace noise with 0 mean.
The parameter $\lambda$ can control the strength of Laplace noise, and larger $\lambda$ can bring better privacy protection.
The function $\text{clip}(x,y)$ is used to limit the value of $x$ with the scale of $y$.
It is motivated by some studies which show that applying gradient clipping can help avoid potential gradient explosion and is beneficial for model training~\cite{zhang2019gradient}.
Denote the randomized gradient as $\widetilde{g}_u=\mathcal{M}(g_u)$.
After clip and randomization operation, it is more difficult to infer the raw user behaviors from the gradients.
The user client uploads the randomized local model gradient $\widetilde{g}_u$ to the server.

In our \textit{FedNewsRec} framework, a server is used to maintain the news recommendation model and update it via the model gradients from a large number of users.
In each round, the server randomly selects a random fraction $r$ (e.g., 10\%) of the user clients, and sends the current news recommendation model $\Theta$ to them.
Then it collects and aggregates the local model gradients from the selected user clients as follows:
\begin{equation}
    \overline{g} = \frac{1}{\sum_{u\in\mathcal{U}}|\mathcal{B}_u|}  \sum_{u\in\mathcal{U}} |\mathcal{B}_u| \cdot \widetilde{g}_u,
     \label{agg_gradient}
\end{equation}
where $\mathcal{U}$ is the set of users selected for the learning process in this round, and $\mathcal{B}_u$ is the set of behaviors of user $u$ for local model gradient computation.

Then the aggregated gradient $\overline{g}$ is used to update the global news recommendation model $\Theta$ maintained in the server:
\begin{equation}\label{eq.update}
    \Theta = \Theta - \eta \cdot \overline{g},
\end{equation}
where $\eta$ is the learning rate.
The updated global model is then distributed to user devices to update their local models.
This process is repeated until the model training converges.







\subsection{Discussions on Privacy Protection}

Next, we discuss why our \textit{FedNewsRec} framework can protect user privacy in news recommendation model training.
First, in our method the private user behavior data is stored on user own devices, and is never uploaded to server.
Only the model gradients inferred from the local user behaviors are communicated with the server.
According to the data processing inequality~\cite{mcmahan2017communication}, these gradients never contain more private information than the raw user behaviors, and usually contain much less information~\cite{mcmahan2017communication}.
Thus, the user privacy can be better protected compared with the centralized storage of user behavior data as did in existing news recommendation methods.
Second, the local model gradients are computed from a group of user behaviors instead of a single behavior.
Thus, it is not very easy to infer a specific behavior from the local model gradients uploaded to server.
Third, we apply the local differential privacy technique to the local model gradients before uploading by adding Laplace noise to them.
It can strengthen the privacy protection of the private information in local model gradients.
According to ~\cite{choi2018guaranteeing}, Laplace noise in LDP can achieve $\epsilon$-local differential privacy, and $\epsilon=\frac{\max_{v,v'} |\mathcal{M}(v)-\mathcal{M}(v')|}{\lambda}$, where $v$ and $v'$ are arbitrary values in local model gradient.
Since the upper bound of $\max_{v,v'} |\mathcal{M}(v)-\mathcal{M}(v')|$ in our \textit{FedNewsRec} framework is $2\delta$, the upper bound of the privacy budget $\epsilon$ is $\frac{2\delta}{\lambda}$.
We can see that by increasing $\lambda$ (i.e., the strength of the noise), we can achieve a smaller privacy budget $\epsilon$ which means better privacy protection.
However, strong noise will hurt the accuracy of aggregated gradients.
Thus, $\lambda$ should be selected based on the trade-off between privacy protection and model performance.

\section{Experiment}

\subsection{Dataset and Experimental Settings}

\bluefont{
Our experiments were conducted on a public news recommendation dataset (named \textit{Adressa}) collected from a Norwegian news website~\cite{gulla2017adressa} and another real-world dataset collected from Microsoft News\footnote{https://www.msn.com/en-us} (named \textit{MSN-News}).\footnote{Our dataset and codes will be publicly available in https://github.com/JulySinceAndrew/FedNewsRec-EMNLP-Findings-2020.}
For the \textit{Adressa} dataset, following~\citet{hu2020graph}, we used user logs in the first five days to construct users' click history, used logs in the 6-th day for model training, and used logs in the 7-th day for model evaluation.
Since the \textit{Adressa} dataset does not contain non-clicked data, we randomly sampled 20 news as negative samples for each click to construct the test set.}
For the \textit{MSN-News} dataset, we randomly sampled 100,000 users and their behavior logs in 5 weeks (from October 19 to November 22, 2019).
We assume that the behavior logs of different users are stored in a decentralized way to simulate the real application of privacy-preserving news recommendation model training.
We used the behaviors in the last week for test and the remaining behaviors for training.
In addition, since in practical applications not all users can participate the model training, we randomly selected half of the users for training and tested the model on all users.
The detailed statistics of the two datasets are listed in Table~\ref{Table_Dataset}.
Following~\cite{wu2019npa}, we used the average scores of AUC, MRR, nDCG@5, nDCG@10 of all impressions in the test set to evaluate the performance.
\bluefont{We repeated each experiment five times and reported average results and stand errors.
}

In experiments we used the 300-dimensional pre-trained Glove embedding to initialize word embeddings.
The number of the self-attention head is 20 and the output dimension of each head is $20$.
The dimension of GRU hidden state is 400.
$H$ in Eq.~(\ref{eq.loss}) is 4.
The fraction $r$ of users participating in model training in each round is 2\%.
The learning rate $\eta$ in Eq.~(\ref{eq.update}) is 0.5.
$\delta$ in Eq.~(\ref{eq.M}) is 0.005 and $\lambda$ in Eq.~(\ref{eq.La}) is 0.015.
These hyper-parameters are all selected according to cross-validation on the training set.

\begin{table}[]

\centering
\resizebox{0.45\textwidth}{!}{

\begin{tabular}{ccc}
\hline
                      & \textit{MSN-News} & \textit{Adressa} \\ \hline
\# users              & 100,000           & 528,514          \\
\# news               & 118,325           & 16,004           \\
\# impressions        & 1,341,853         & -                \\
\# positive behaviors & 2,006,289         & 2,411,187        \\
\# negative behaviors & 48,051,601        & -                \\
avg. \# title length  & 11.52             & 6.60             \\ \hline
\end{tabular}

}
\caption{The statistical information of the dataset.}
\vspace{-0.1in}
\label{Table_Dataset}
\end{table}

\subsection{Effectiveness Evaluation}

\begin{table*}[h]

\centering

\resizebox{0.98\textwidth}{!}{
\begin{tabular}{c|cccc|cccc}
\hline
\multirow{2}{*}{Method} & \multicolumn{4}{c|}{MSN-News}                                      & \multicolumn{4}{c}{Adressa}                                       \\ \cline{2-9} 
                        & AUC            & MRR            & nDCG@5         & nDCG@10         & AUC            & MRR            & nDCG@5         & nDCG@10        \\ \hline
FM                      & 58.41$\pm$0.04 & 27.19$\pm$0.05 & 28.98$\pm$0.04 & 34.57$\pm$0.06  & 61.94$\pm$0.80 & 26.59$\pm$0.33 & 22.69$\pm$0.54 & 32.17$\pm$0.46 \\
DFM                     & 61.25$\pm$0.26 & 28.68$\pm$0.10 & 30.62$\pm$0.21 & 36.38$\pm$0.23  & 65.14$\pm$0.69 & 34.74$\pm$0.89 & 33.17$\pm$1.46 & 39.79$\pm$1.08 \\ \hline
EBNR                    & 63.64$\pm$0.15 & 29.50$\pm$0.14 & 31.57$\pm$0.13 & 37.38$\pm$0.20  & 65.70$\pm$0.72 & 30.23$\pm$0.49 & 29.37$\pm$0.53 & 36.38$\pm$0.44 \\
DKN                     & 62.38$\pm$0.19 & 29.40$\pm$0.15 & 31.59$\pm$0.11 & 37.27$\pm$0.21  & 67.53$\pm$1.90 & 32.33$\pm$2.79 & 31.84$\pm$2.78 & 39.96$\pm$2.52 \\
DAN                     & 62.54$\pm$0.23 & 29.44$\pm$0.18 & 31.67$\pm$0.14 & 37.31$\pm$0.25  & 64.03$\pm$3.10 & 33.37$\pm$2.63 & 31.61$\pm$3.03 & 38.60$\pm$3.02 \\
NAML                    & 64.52$\pm$0.24 & 30.93$\pm$0.17 & 33.39$\pm$0.16 & 39.07$\pm$0.19  & 69.20$\pm$2.07 & 35.18$\pm$1.49 & 34.78$\pm$1.85 & 42.34$\pm$1.97 \\
NPA                     & 64.29$\pm$0.20 & 30.63$\pm$0.15 & 33.11$\pm$0.17 & 38.89$\pm$0.23  & 66.70$\pm$2.42 & 34.68$\pm$1.77 & 33.72$\pm$2.09 & 41.18$\pm$1.99 \\
NRMS                    & 65.72$\pm$0.16 & 31.85$\pm$0.20 & 34.59$\pm$0.18 & 40.25$\pm$0.17  & 67.97$\pm$2.23 & 33.16$\pm$2.54 & 32.37$\pm$3.59 & 40.41$\pm$2.82 \\ \hline
FCF                     & 51.03$\pm$0.27 & 22.24$\pm$0.14 & 22.97$\pm$0.21 & 28.44$\pm$0.23  & 53.33$\pm$1.28 & 23.04$\pm$2.68 & 20.24$\pm$2.77 & 27.09$\pm$2.61 \\ \hline
FedNewsRec              & 64.65$\pm$0.15 & 30.60$\pm$0.09 & 33.03$\pm$0.11 & 38.77$\pm$0.10  & 69.91$\pm$2.53 & 35.55$\pm$1.85 & 33.74$\pm$2.45 & 41.47$\pm$2.78 \\ \hline \hline
CenNewsRec              & 66.45$\pm$0.17 & 31.91$\pm$0.22 & 34.62$\pm$0.18 & 40.33$\pm$ 0.24 & 71.02$\pm$2.09 & 36.31$\pm$2.52 & 35.73$\pm$3.71 & 43.98$\pm$2.52 \\ \hline
\end{tabular}
}

\caption{The news recommendation results of different methods.}

\label{Table_Mainexp}
\end{table*}

First, we verify the effectiveness of the proposed \textit{FedNewsRec} method.
We compared with many methods, including:
(1) \textit{FM}~\cite{rendle2012factorization}, factorization machine, a classic method for recommendation;
(2) \textit{DFM}~\cite{lian2018towards}, deep fusion model for news recommendation;
(3) \textit{EBNR}~\cite{okura2017embedding}, using GRU for user modeling~\cite{cho2014properties};
(4) \textit{DKN}~\cite{wang2018dkn}, using knowledge-aware CNN network for news representation in news recommendation;
(5) \textit{DAN}~\cite{danzhu2019}, using CNN to learn news representations from both news title and entities and using LSTM to learn user representations;
(6) \textit{NAML}~\cite{wu2019ijcai}, learning news representations via attentive multi-view learning;
(7) \textit{NPA}~\cite{wu2019npa}, using personalized attention network to learn news and user representations;
(8) \textit{NRMS}~\cite{wu2019neuralc}, learning representations of news and users via multi-head self-attention network;
(9) \textit{FCF}~\cite{ammad2019federated}, a federated collaborative filtering method for recommendation;
(10) \textit{CenNewsRec}, which has the same news recommendation model with \textit{FedNewsRec} but is trained on centralized user behavior data. 


The results are listed in Table~\ref{Table_Mainexp}.
First, by comparing \textit{FedNewsRec} with SOTA news recommendation methods such as \textit{NRMS}, \textit{NPA} and \textit{EBNR}, our method can achieve comparable and even better performance on news recommendation.
It validates the effectiveness of our approach in learning accurate models for personalized news recommendation.
Moreover, different from these existing news recommendation methods which are all trained on centralized storage of user behavior data, in our \textit{FedNewsRec} the user behavior data is stored on local user devices and is never uploaded.
Thus, our method can train accurate news recommendation model and meanwhile better protect user privacy.

Second, our method can perform better than existing federated learning based recommendation methods like \textit{FCF}~\cite{ammad2019federated}.
The performance of \textit{FCF} is not good in news recommendation.
This is because \textit{FCF} requires each user and each item to participate the training process to learn their embeddings.
However, in practical application not all the users can participate the training due to different reasons.
In addition, news articles on online news platforms expire very quickly, and new news articles continuously emerge.
Thus, many items for recommendation are news items, and unseen in the training data, which cannot be handled by \textit{FCF}.
In our method we learn news representations from news content and learn user representations from their behaviors using neural models.
Therefore, our method can handle the problem of new users and new items, and is more suitable for news recommendation scenario.

Third, \textit{FedNewsRec} performs worse than \textit{CenNewsRec} which has the same news recommendation model with \textit{FedNewsRec} but is trained on the centralized user behavior data.
This is intuitive since centralized data is more beneficial for model training than decentralized data.
In addition, in \textit{FedNewsRec} we apply local differential privacy technique with Laplace noise to protect the private information in model gradients, which leads to the aggregated gradient for model update less accurate.
Luckily, the gap between the performance of \textit{FedNewsRec} and \textit{CenNewsRec} is not very big.
Thus, our \textit{FedNewsRec} method can achieve much better privacy protection at the cost of acceptable performance decline. 
These results validate the effectiveness of our method.


\subsection{Influence of User Number}

\begin{figure}
    \centering
    \resizebox{0.48\textwidth}{!}{
    \includegraphics{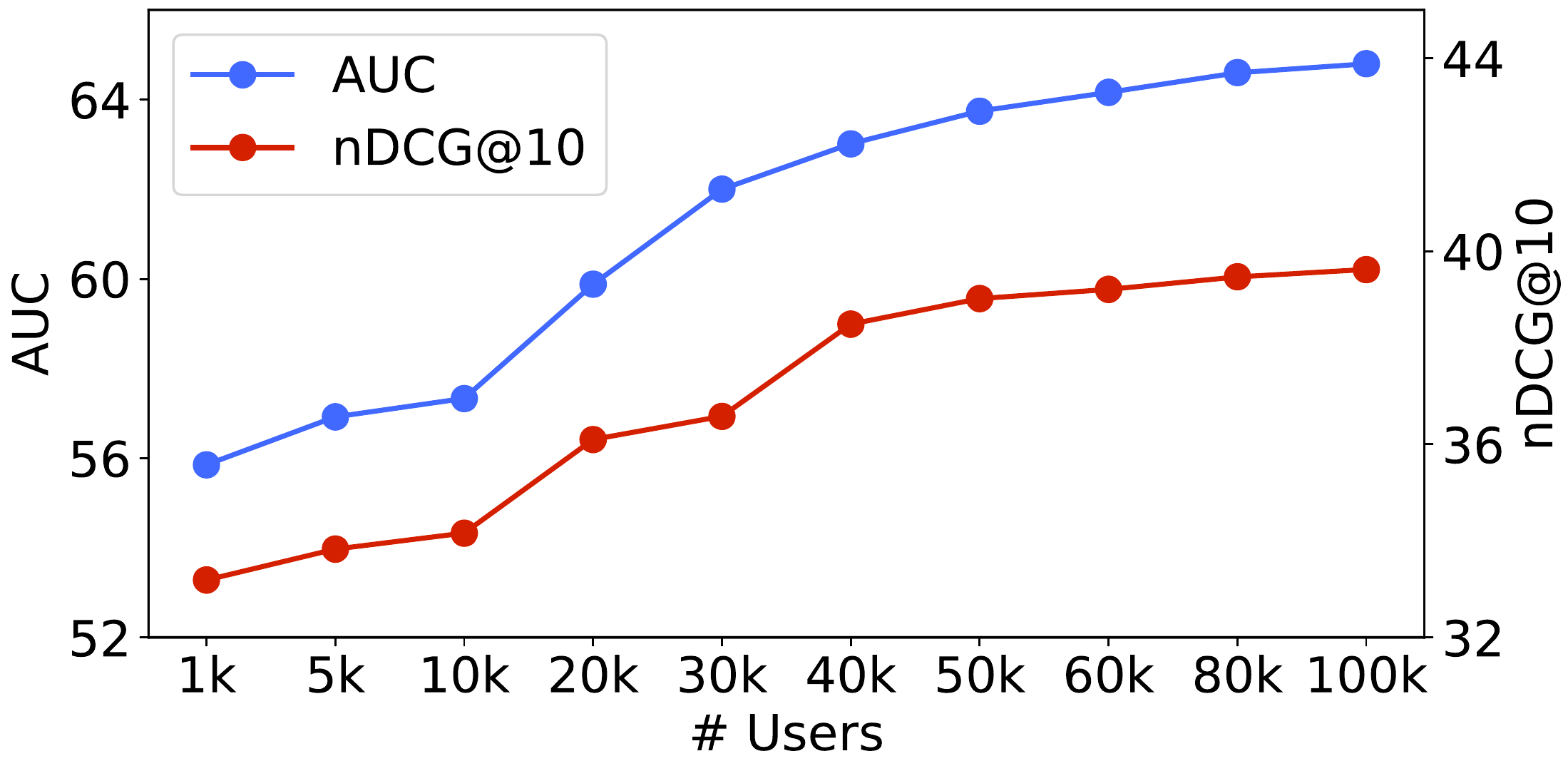}
    }

    \caption{Performance with different numbers of users.}

    \label{fig_user}
\end{figure}

In this section, we explore whether our \textit{FedNewsRec} method can exploit the useful behavior information of massive users in a federated way to train accurate news recommendation models.
\bluefont{In the following sections, we only show the experimental results on the \textit{MSN-News} dataset.}
We randomly select different numbers of users for model training, and use all users for evaluation.
The experimental results are shown in Fig.~\ref{fig_user}.

From Fig.~\ref{fig_user} we have several observations.
First, when the number of users is small (e.g., 1000), the performance of news recommendation model trained on the behavior data of these users is not satisfactory.
This is because the behaviors of a single user are usually very limited, and behavior data of a small number of users is insufficient to train accurate news recommendation model.
This result validates the motivation of our \textit{FedNewsRec} method to coordinate a large number of users in a federated way for model training.
Second, when the number of users participating in training increases, the performance of \textit{FedNewsRec} improves.
It indicates that \textit{FedNewsRec} can effectively exploit the useful behavior information from different users to collectively train an accurate news recommendation model, which validates the effectiveness of our framework.
Third, when the number of users is big enough, further incorporating more users can only bring marginal performance improvement.
This result shows that a reasonable number of users are sufficient for news recommendation model training, and it is unnecessary to involve too many or all users which is costly and impractical.

\subsection{Hyper-parameter Analysis}

\begin{figure}
	\centering

	\subfigure[Model performance.]{
			\includegraphics[width=0.46\linewidth]{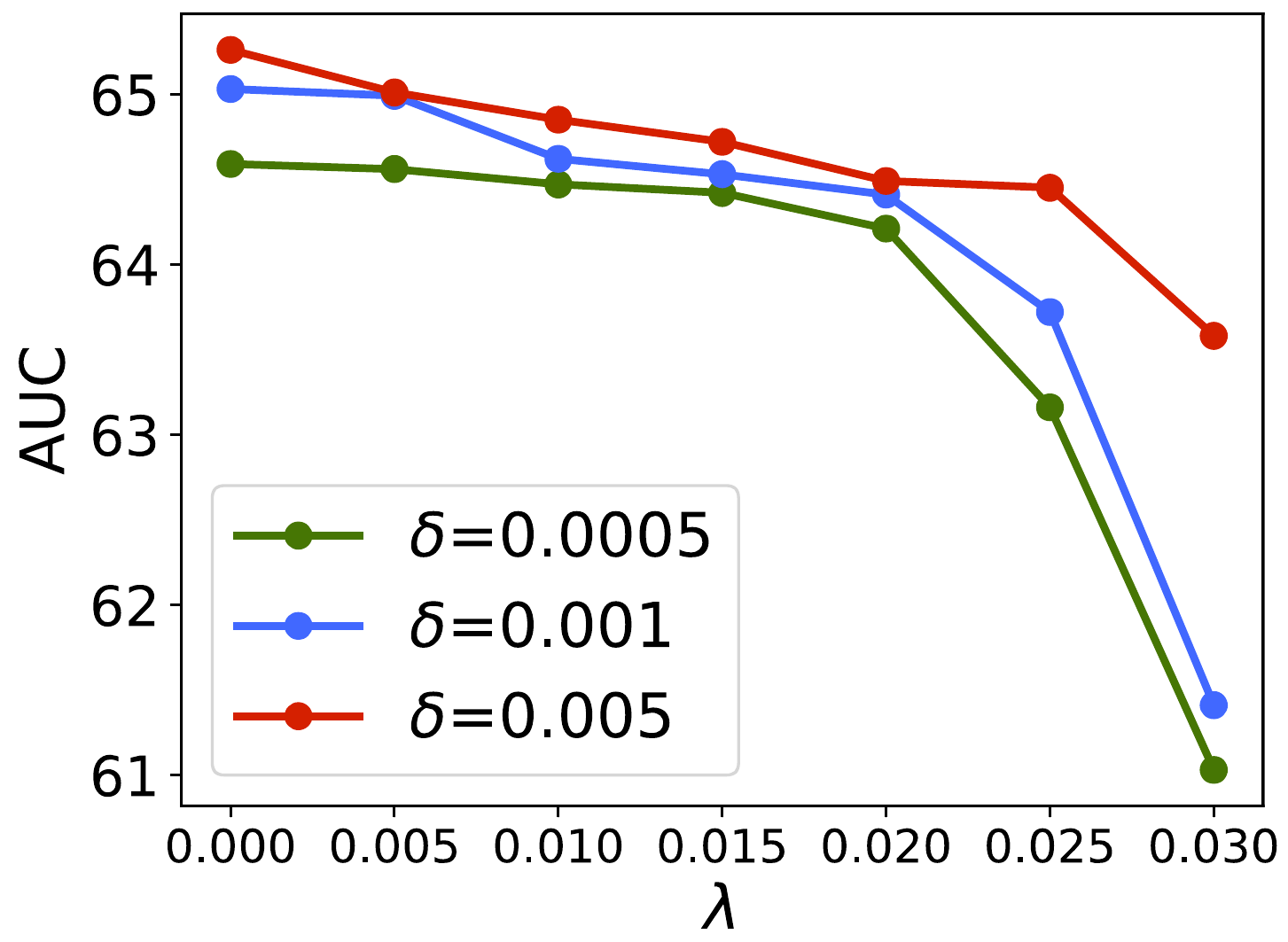}\label{fig_para_performance}
 		}
	\subfigure[Privacy budget $\epsilon$.]{
			\includegraphics[width=0.46\linewidth]{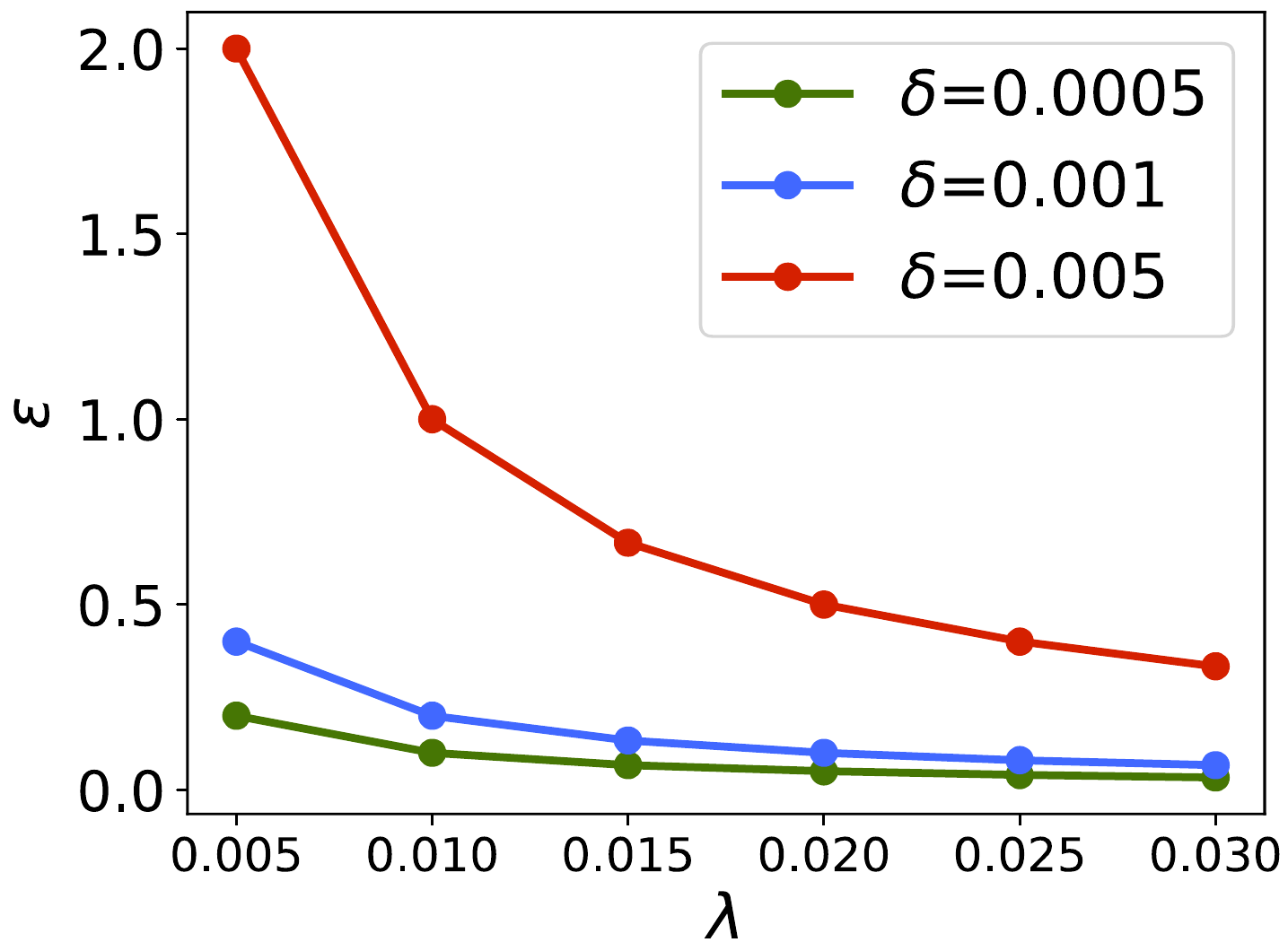}\label{fig_para_privacy}
		}

	\caption{Influence of the hyper-parameters $\lambda$ and $\delta$.}

	\label{fig_para}
\end{figure}

In this section, we explore the influence of hyper-parameters on our method.
We show the results of two important hyper-parameters, i.e., $\delta$ in Eq.~(\ref{eq.M}) and $\lambda$ in Eq.~(\ref{eq.La}) which serve in the local differential privacy module of our \textit{FedNewsRec} framework.
The results are shown in Fig.~\ref{fig_para}.
In Fig.~\ref{fig_para_performance} we show the performance of our method with different $\lambda$ and $\delta$ values.
We find that a large $\lambda$ value can lead to the performance decline.
This is because larger $\lambda$ means stronger Laplace noise added to the gradients in LDP module, making the aggregated gradient for model update less accurate.
In addition, our method tends to have better performance when $\delta$ is larger.
This is because fewer gradients will be affected in the clip operation when $\delta$ is larger.
In Fig.~\ref{fig_para_privacy} we show the upper bound of the privacy budget, i.e., $\epsilon$ in Section 3.3, with different $\lambda$ and $\delta$ values.
We can find that with larger $\lambda$ value and smaller $\delta$ value, the privacy budget $\epsilon$ becomes lower, which means better privacy protection.
This is intuitive, since larger $\lambda$ value and smaller $\delta$ value indicate that stronger noise is added and more gradient values are clipped, making it more difficult to recover the original model gradients.
Combining Fig.~\ref{fig_para_performance} and Fig.~\ref{fig_para_privacy} we can see that the better privacy protection is achieved by some sacrifice of the performance, and we need to select $\lambda$ and $\delta$ values based on the trade-off between privacy protection and news recommendation performance.

\subsection{Convergence Analysis}
\begin{figure}
    \centering
    \resizebox{0.4\textwidth}{!}{
    \includegraphics{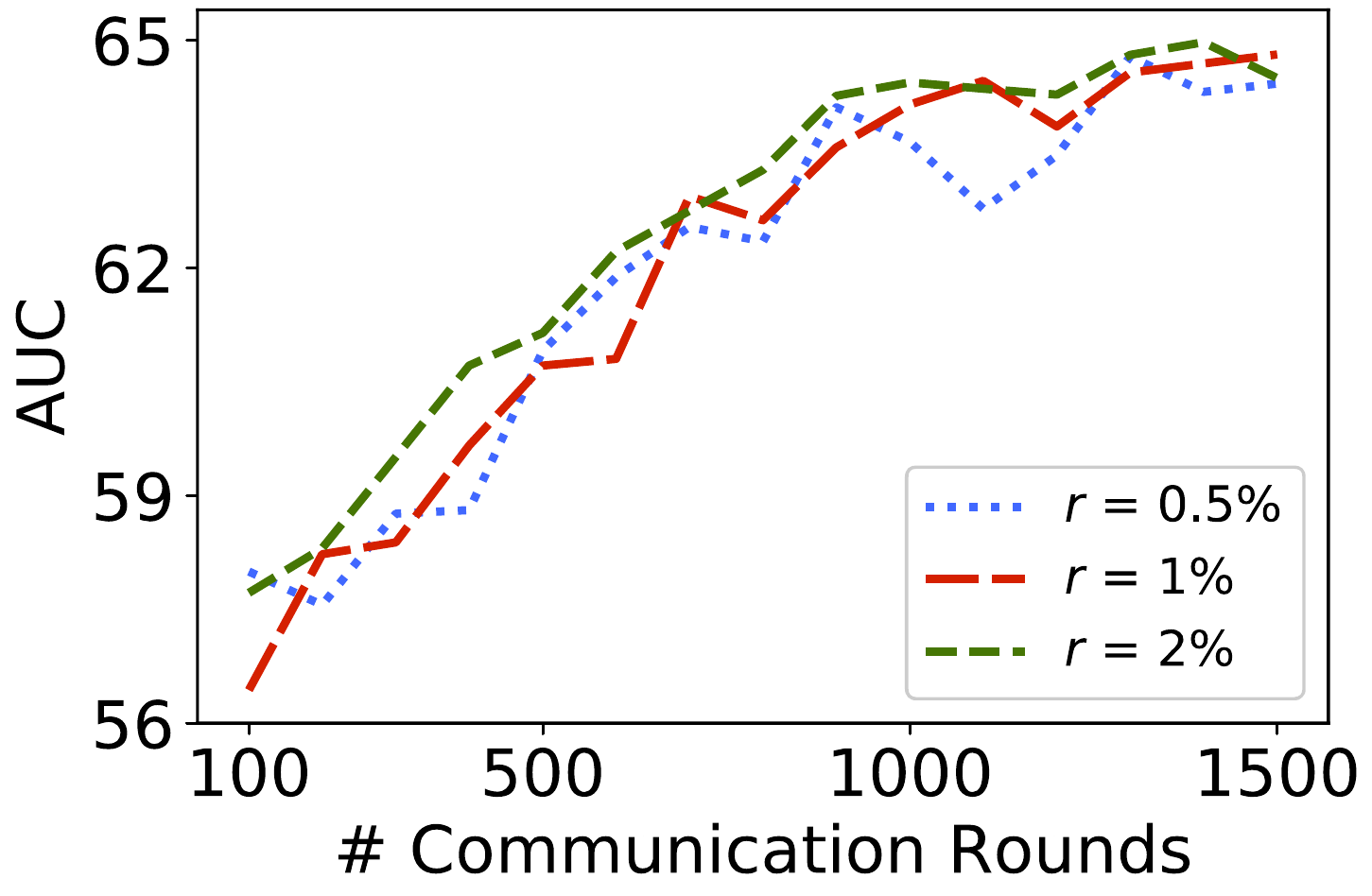}
    }

    \caption{Convergence of model training.}

    \label{fig_ratio}
\end{figure}

Next we explore the convergence of the model training in \textit{FedNewsRec}, and the results are shown in Fig.~\ref{fig_ratio}.
We can see that the training process can converge in about 1,500 rounds under different settings of $r$ (i.e., ratio of selected users for model training in each round).
It indicates that \textit{FedNewsRec} can train news recommendation model efficiently.

\subsection{Effectiveness of User Model}

\begin{figure}
    \centering
    \resizebox{0.45\textwidth}{!}{
    \includegraphics{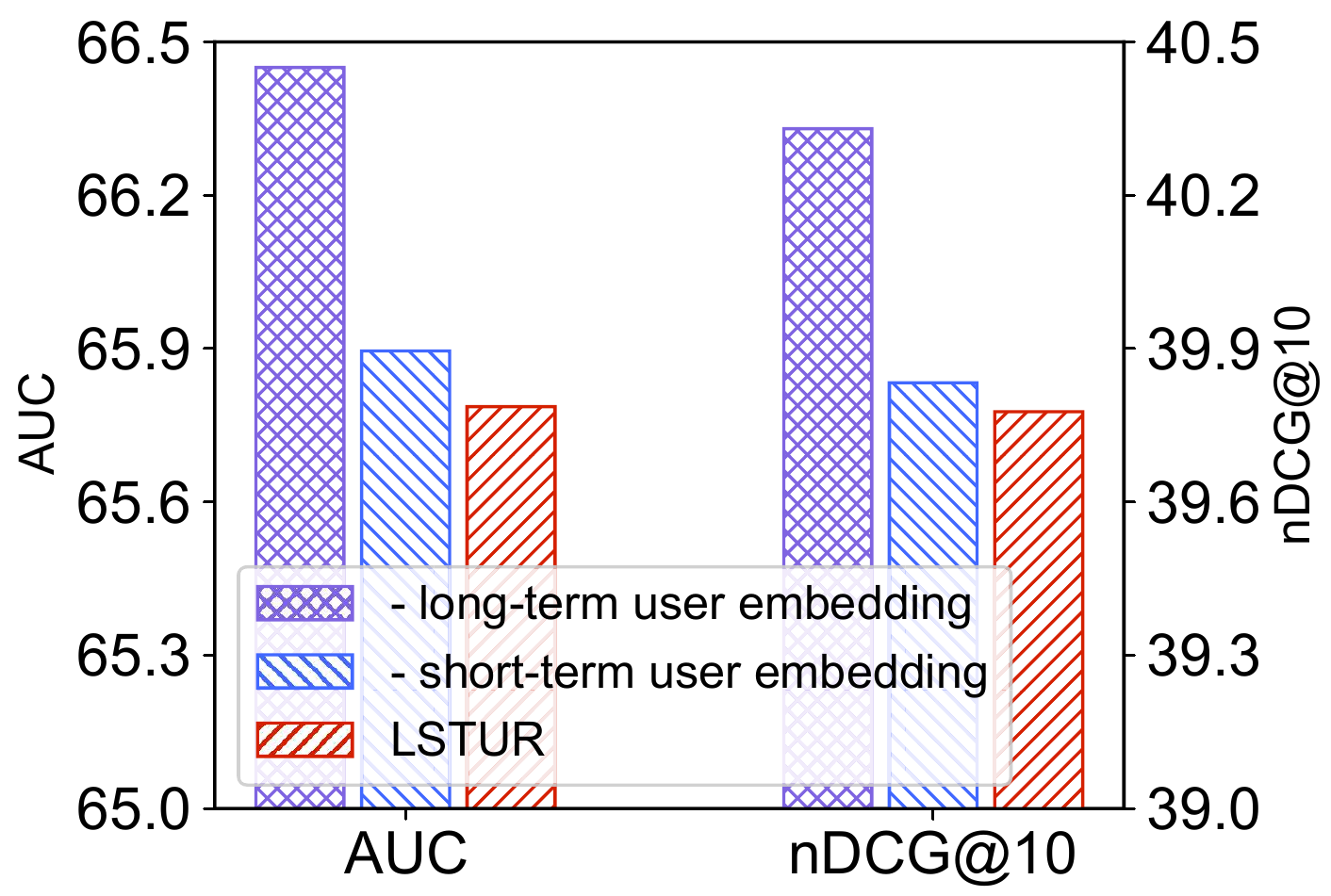}
    }

    \caption{Effectiveness of User Model}

    \label{fig_ue}
\end{figure}

\bluefont{
In this section, we conduct ablation studies to evaluate the effectiveness of the short- and long- term user interest modeling in our user model.
The experimental results are shown in Fig.~\ref{fig_ue}, from which we have several observations.
First, after removing the short-term user embedding, the performance of our method declines.
This is because users sometimes tend to read news related to the topics they recently cared about~\cite{an2019neural}.
Our user model learns the short-term user embedding from the sequence of users' recent clicked news via a GRU network, which can effectively capture users' short-term interest.
Thus, removing the short-term user embedding makes the unified user embedding loss some information of users' recent reading preference and causes performance decline.
Second, after removing the long-term user embedding, the performance of our method also declines.
This is because users may read some news according to their long-term interests, which may not be reflected by their recent reading history~\cite{an2019neural}.
To address this issue, our user model learns long-term user embedding by capturing the relatedness among users' clicked news, which can effectively capture users' long-term interest.
After removing it, the unified user embedding losses the information of the long-term interest, which hurts the recommendation accuracy.
}

\section{Conclusion}

In this paper, we propose a privacy-preserving method for news recommendation model training.
Different from existing methods which rely on centralized storage of user behavior data, in our method the user behaviors are locally stored on user devices.
We propose a \textit{FedNewsRec} framework to coordinate a large number of users to collectively train accurate news recommendation models from the behavior data of these users without the need to upload it.
In our method each user client computes local model gradients based on the user behaviors on device, and sends them to server.
The server coordinates the training process and maintains a global news recommendation model.
It aggregates the local model gradients from massive users and updates the global model using the aggregated gradient.
Then the server sends the updated model to user clients and this process is repeated for multiple rounds.
In order to further protect the private information in the local model gradients, we apply local differential privacy to them by adding Laplace noise.
The experiments on real-world dataset show that our method can achieve comparable performance with SOTA news recommendation methods, and meanwhile can better protect user privacy.

\section{ACKNOWLEDGMENTS}
This work was supported by the National Key Research and Development Program of China under Grant number 2018YFC1604002, and the National Natural Science Foundation of China under Grant numbers U1836204, U1705261, and 61862002.

\bibliography{anthology}
\bibliographystyle{acl_natbib}

\end{document}